\documentclass[aps,pre,twocolumn,showpacs,10pt,longbibliography]{revtex4-1}
\usepackage{graphicx}
\usepackage{epstopdf}
\usepackage{amsmath}
\usepackage{amssymb}
\usepackage{default}
\newcommand{\be}{\begin{equation}}
\newcommand{\bea}{\begin{eqnarray}}
\newcommand{\bc}{\begin{center}}            
\newcommand{\ee}{\end{equation}}
\newcommand{\eea}{\end{eqnarray}}
\newcommand{\ec}{\end{center}}

\newcommand{\baa}{\begin{eqnarray*}}
\newcommand{\eaa}{\end{eqnarray*}}
\begin{document}
\title{Performance optimization 
of low-dissipation thermal machines revisited}
\author{Ramandeep S. Johal} 
\email[e-mail: ]{rsjohal@iisermohali.ac.in}
\altaffiliation[Permanent address: ]{Indian Institute of Science Education and Research Mohali,
Department of Physical Sciences, 
Sector 81, S.A.S. Nagar, Manauli PO 140306, Punjab, India}
\affiliation{Max-Planck-Institut f\"{u}r Physik komplexer Systeme,  N\"{o}thnitzerstra{\ss}e 38,   01187 Dresden, Germany}
\begin{abstract}
We revisit the optimization of performance of finite-time Carnot  
machines satisfying the low-dissipation assumption.
The standard procedure seeks to optimize an objective
function, such as power output of the engine,
over the durations of contacts between the 
working medium and the heat reservoirs. 
This procedure may lead to unwieldy equations 
at the optimum of some objective functions. We propose an alternate 
scheme in which the output or input work is first optimized 
for a given cycle time, followed by an optimization
 of another objective function over the cycle time.
This optimization problem is solved 
in a much simplified manner, with closed-form
expressions for figures of merit. 
The approach is demonstrated for various
objective functions, both for engines as well 
as refrigerators.
\end{abstract}
\maketitle
\section{Introduction}
Optimization of performance of finite-time thermal machines 
has been intensely studied for many years now
\cite{Berry1984, Bejan1996, Salamon2001, Andresen2011, Seifert_2012}.
In recent years, the low-dissipation model
has been proposed and applied to heat engines and refrigerators
with presumably large cycle times and so, close to the reversible
limit. The low-dissipation regime is characterized by 
the following dependence: the entropy 
generated in a heat-exchange process is
 inversely proportional to the duration of the process. 
It was initially 
 derived for a mesoscopic, brownian heat engine
treated within stochastic thermodynamic framework \cite{Schmiedl2008},
and was later adapted for  finite-time macroscopic engines \cite{Esposito2010}. 
It is observed at the optimal performance of quantum dot Carnot 
engine based on the master equation approach \cite{Lindenberg2010}, 
and within a perturbative approach for slowly driven 
open quantum systems \cite{Cavina2017}.

Because of its simplicity, the low-dissipation model has attracted a lot 
of attention \cite{Roco2012, Roco2012b, Izumida_2012,  
Broeck2013, Holubec2016, Holubec2016e,Ryabov2016, Roco2017, Johal2017}. Furthermore, 
there is no explicit requirement on the form of heat-transfer law, or  
the temperature difference between
the heat reservoirs to be small, unlike in endoreversible models \cite{Esposito2010}. 
Still, the optimization
problem may become cumbersome, or even intractable, with some objective functions.
In this paper, we propose an alternate two-step optimization scheme
 which yields the optimal solution 
in a quite simplified manner, while predicting the essential characteristics
of the model, such as
closed-form
expressions for figures of merit as well as 
the bounds satisfied by them within the domain of applicability
of the model. The utility
of the approach is demonstrated on various objective functions. 

The plan of the paper is as follows. In Section II, we briefly describe
the basic features low-dissipation Carnot engine. In Section III, 
we first optimize the work output for a given cycle time and 
discuss its main features.
In Section IV, we optimize other objective functions for the engine,
giving explicit expressions
for the efficiency. In Section V, we treat the model
of a refrigerator and derive expressions for the 
coefficient of performance at optimum of different objective functions.
Sections VI is devoted to a discussion of some  novel features
of the work-optimized model and Section VII contains
the conclusions.
\section{Low-dissipation model}
Consider a two heat-reservoirs set up, with hot ($h$) and cold 
($c$) temperatures, $T_h$ and $T_c$. A heat engine runs through a four-step 
cycle by coupling to these reservoirs alternately.
The cycle consists of two thermal contacts lasting for 
time intervals $\tau_h$ and $\tau_c$, and two adiabatic steps whose time 
intervals are considered negligible in comparison to the other time scales.
Now, the change in entropy of the working medium 
during heat transfer at the hot/cold  
contact, can be split as: $\Delta S_{j} = \Delta_{\rm rev} S_{j} + \Delta_{\rm ir} S_{j}$,
with $j=h,c$. Here, the first term accounts for a reversible 
heat transfer, whereas the second term denotes an irreversible entropy 
generation during the process. Now,  
the low-dissipation behavior is quantified as: 
$T_j \Delta_{\rm ir} S_{j} = \sigma_j/ t_j + O(1/ t_{j}^{2})$,
where $\sigma_{j}$ is the dissipation constant \cite{Esposito2010,Broeck2013},
and the higher order terms are considered neglegible due to the large 
durations. Thus
at the hot and the cold contact, we respectively have
\bea
\Delta S_h &=& \frac{Q_h}{T_h} + \frac{\sigma_h}{T_h t_h},
\label{dsh} \\
\Delta S_c &=& -\frac{Q_c}{T_c} + \frac{\sigma_c}{T_c t_c},
\label{dsc}
\eea
where $Q_j > 0$. Given that the other two steps in the heat cycle are 
adiabatic---with no entropy changes---the cyclic process 
within the working medium implies 
$\Delta S_h + \Delta S_c = 0$. In other words, 
$\Delta S_h = -\Delta S_c = \Delta S > 0$,
where the value $\Delta S$ is preassigned.
Then the amount of heat exchanged with each reservoir can be written as:
\bea
Q_h &=& T_h \Delta S - \frac{\sigma_h}{t_h}, 
\label{qh} \\
Q_c &=& T_c  \Delta S + \frac{\sigma_c}{t_c}. 
\label{qc}
\eea
The work extracted in a cycle with the time period $t \approx t_h + t_c$ is,  
$W = Q_h - Q_c$, given by  
\be
W (t_h, t_c) = \Delta T \Delta S -  \frac{\sigma_h}{t_h} - \frac{\sigma_c}{t_c},
\label{whc}
\ee
where $\Delta T = T_h - T_c$.
Now, in the standard optimzations of the LD model, the parameter
$\Delta S$ is held fixed during variations of the time intervals.
Clearly, for a given $\Delta S$,  as each $t_j \to \infty$,
the work approaches its maximum value of $\Delta T \Delta S \equiv W_{\rm rev}$,
which is referred to as the reversible work, under conditions of
a fixed $\Delta S$. Therefore, the difference $W_{\rm rev}-W$
represents the lost work due to entropy production. 
Recently, we showed that 
each loss term above can be obtained from a linear-irreversible 
engine running between an infinite heat reservoir and a finite
heat sink or source for a given time ($t_h$ or $t_c$) \cite{iyy2019}.  
\section{Optimal work for a given cycle time}
Now, instead of choosing $t_h$ and $t_c$ as the control parameters 
that may be tuned in order to optimize the overall performance 
of the engine \cite{Schmiedl2008, Esposito2010, Broeck2013}, 
let us  define $t_h$ and $t_c$ in terms of
the fraction of the total cycle time as:  
$t_h = \gamma t$, and $t_c = (1-\gamma) t$,
obtaining 
\be
W (\gamma, t) = \Delta T \Delta S -  
\left(\frac{\sigma_h}{\gamma} + \frac{\sigma_c}{1-\gamma} \right) \frac{1}{t}.
\label{wgte2}
\ee
As a first step towards optimization of the engine's performance, we 
maximize the irreversible work $W$ 
for a fixed value of the time interval $t$. This amounts to tuning the 
parameter $\gamma$. 
Thus, setting:
\be
\left. \frac{\partial W}{ \partial \gamma} \right \vert_{t,\Delta S}  = 0,
\ee
we obtain the optimum value of $\gamma$ as
\be
\hat{\gamma} = \frac{\sqrt{\sigma_h}}{\sqrt{\sigma_h} + \sqrt{\sigma_c}},
\label{gcap}
\ee
which is function only of the ratio of the dissipation constants.
In the following, we seek to optimize the sub-class of low-dissipation
models which operate at optimal work
in a given time.
Thus, in our model, the (maximum) work output for a cycle of time $t$ is:
\be
\hat{W}(t) = \Delta T \Delta S - \frac{({\sqrt{\sigma_h} + \sqrt{\sigma_c}})^2}{t},
\label{wmaxt}
\ee
which is equivalent to finite-time availability \cite{Tolman1948, Andresen1983}
for the low-dissipation engine. 

Then, the heat absorbed from the hot reservoir is:
\be
\hat{Q}_h (t) = T_h \Delta S - \frac{\sqrt{\sigma_h}({\sqrt{\sigma_h} + \sqrt{\sigma_c}})}{t}. 
\label{qh2}
\ee
Thus, the efficiency under these conditions is: $ \hat{\eta} (t) = \hat{W}/\hat{Q}_h$.

Similarly, the heat rejected to the cold reservoir is:
\be
\hat{Q}_c (t) = T_c \Delta S + \frac{\sqrt{\sigma_c}({\sqrt{\sigma_h} + \sqrt{\sigma_c}})}{t}. 
\label{qh3}
\ee
\par
We notice two limiting cases here. For a finite value
of $\sigma_h$, if we have $\sigma_c \ll \sigma_h$, which is
equivalent to the condition $\hat{\gamma} \approx 1$, Eqs.
(\ref{wmaxt})-(\ref{qh3}) simplify to: $\hat{W}(t) \approx \Delta T \Delta S
-\sigma_h /t$, $\hat{Q}_h (t) \approx T_h \Delta S - {\sigma_h}/t$, and 
$\hat{Q}_c \approx T_c \Delta S$. In other words, the heat exchange
at the cold end approaches its reversible value, in this limit. 
On the other hand, for a given finite value
of $\sigma_c$, if we have $\sigma_h \ll \sigma_c$, which is
equivalent to the condition $\hat{\gamma} \approx 0$, Eqs.
(\ref{wmaxt})-(\ref{qh3}) simplify to: $\hat{W}(t) \approx \Delta T \Delta S
-\sigma_c /t$, $\hat{Q}_h \approx T_h \Delta S$, and 
$\hat{Q}_c (t) \approx T_c \Delta S + {\sigma_c}/t$. 
Thus, when the strength of dissipation at the 
hot end is negligible as compared to the cold end, then, at optimal work, 
the heat exchange
at the hot end can be approximated to be reversible. 

Further, upon eliminating time 
$t$ from Eqs. (\ref{qh2}) and (\ref{qh3}), we obtain the following 
interesting equality:
\be 
\frac{\hat{\gamma} \hat{Q}_c + (1-\hat{\gamma}) \hat{Q}_h}
{\hat{\gamma} T_c + (1-\hat{\gamma}) T_h}
= \Delta S.
\label{dseq}
\ee
This also makes it clear that as $\hat{\gamma} \to 1$, $\hat{Q}_c \to T_c \Delta S$. 
Further, this limit implies $\hat{t}_h/t \to 1$, while $\hat{t}_c/t \to 0$. 
Thus, for a smaller dissipation at a thermal contact, one has
to spend a smaller fraction of the total given time for that
process. 
Similarly, as $\hat{\gamma} \to 0$, $\hat{Q}_h \to T_h \Delta S$,
and analogous conclusions can be drawn.

Equivalently, we may consider the ratio of dissipations at
the hot to cold contacts, as follows:
\be
\frac{T_h \Delta_{\rm ir} \hat{S}_h}{T_c \Delta_{\rm ir} \hat{S}_c} =
\frac{\hat{\gamma}}{1-\hat{\gamma}} = \sqrt{\frac{\sigma_h}{\sigma_c}}.
\ee
In this sense, $\hat{\gamma} \to 0$ can be regarded as the limit
in which the dissipation at the hot end becomes negligible {\it in comparison}
to the dissipation at the cold end, and so on. A related fact
is that the average rate of dissipation at the hot and the cold end become
equal at optimum work:
\be
\frac{T_h \Delta_{\rm ir} \hat{S}_h}{\hat{t}_h} = \frac{T_c 
\Delta_{\rm ir} \hat{S}_c}{\hat{t}_c} = 
\frac{({\sqrt{\sigma_h} + \sqrt{\sigma_c}})^2}{t^2}.
\ee
%
\section{Optimization: Second step}
The optimal work derived above, Eq. (\ref{wmaxt}), is still a function of 
the chosen cycle time $t$. An appropriate value of 
this cycle time may be selected as the one which optimizes another 
chosen objective function. In the following, we show 
that, for a variety of objective functions---popular
in the study of finite-time thermodynamics---it is
relatively easy to perform this optimization and thus
to find an optimal cycle time. Interestingly, this 
procedure also yields a closed-form expression for 
the corresponding figure of merit, along with its
lower and upper bounds set by the allowed parameter range.
We show the utility of this approach for low-dissipation 
engines as well as refrigerators. 
\subsection{Power output} 
After knowing the optimal work as a function of the cycle 
time, we may like to extract this work at the fastest rate.
An appropriate objective function to optimize  is then the average 
power output, defined as 
\be
P \equiv \frac{\hat{W}(t)}{t} =  \frac{\Delta T \Delta S}{t} - 
\frac{({\sqrt{\sigma_h} + \sqrt{\sigma_c}})^2}{t^2}.
\label{pwr}
\ee
Note that the power output is defined relative to the optimal 
work in time $t$.
Then, $t^*$, corresponding to    
the maximum of this power, is obtained by 
setting $\partial P/\partial t=0$, which yields 
\be
t^* = \frac{2 ({\sqrt{\sigma_h} + \sqrt{\sigma_c}})^2}
{\Delta T \Delta S},
\label{tstar}
\ee
with the optimal allocation of times for the thermal contacts:
$t_h = \hat{\gamma} t^*$, and $t_c = (1-\hat{\gamma}) t^*$.
The optimal amounts of heat and work are: 
\bea
{Q}_{h}^{*} &=& \left[ T_h -  \frac{\hat{\gamma}}{2} \Delta T \right] \Delta S, \\ 
W^* &=& \frac{\Delta T \Delta S}{2}, \label{wstar}
\label{qhstar}
\eea
from which the efficiency at maximum power,
${\eta}^* ={W}^*/{Q}_{h}^{*}$, follows in the well-known form \cite{Chen1989, Schmiedl2008, Broeck2013}:
\be
{\eta}^* = \frac{\eta_{\rm C}}{2-\hat{\gamma} \eta_{\rm C}}.
\label{etg}
\ee
Note that the same optimum for power may also be obtained by performing 
optimization simultaneously over the pair of variables $t_h$ and $t_c$ \cite{Broeck2013},
which is the standard approach in literature. However, this approach
often becomes involved and an analytic solution
becomes hard to obtain with other objective functions, in general.
In the following, we highlight 
the utility of the present two-step optimization
approach, for the case of engines as well as refrigerators.
\subsection{Per-unit-time efficiency}
First proposed by Ma \cite{Ma}, this objective function
was optimized for the endoreversible model 
in Ref. \cite{Velas1998}. Our first step is
to optimize the work output
for a given time $t$, as described above,  and calculate the 
efficiency at this optimal work, denoted by $\hat{\eta} (t)$.
As the second step, we optimize the function:
\be
\dot{\hat{\eta}} \equiv \frac{\hat{\eta}(t)}{t},
\ee
w.r.t. time $t$.
The solution can be easily worked out and the efficiency
at optimal $\dot{\hat{\eta}}$ is given by:
\be
\hat{\eta}^* = \frac{1}{\hat{\gamma}} \left( 1-\sqrt{1-\hat{\gamma}\eta_{\rm C}} \right),
\label{estar2}
\ee
which is bounded as:  $\eta_{\rm C}/2 \leqslant \hat{\eta}^*
\leqslant 1-\sqrt{1- \eta_{\rm C}}$. Thus the results 
from the endoreversible model \cite{Velas1998} are derived
within the low-dissipation model too, in a simple manner.
\subsection{Efficient power}
An objective function, defined  as the product
of efficiency of the engine and its power output \cite{Stucki},
was optimized for the low-dissipation model
with the standard optimization \cite{VJ2018}, but the solution
turns out to be highly involved. In the present approach, at 
optimal work for the given cycle time $t$, the efficient
power is defined as:
\be
\hat{P}_\eta (t) = \hat{\eta}(t) \frac{\hat{W}(t)}{t}.
\ee
The optimum  of the above function ($\partial \hat{P}_\eta/\partial t =0$)
is easily evaluated by just solving a quadratic equation in $t$. Finally,
the efficiency at optimal efficient power is obtained in
a simple closed form:
\be
\eta^* = \frac{1}{2\hat{\gamma}}
\left[3-\sqrt{9-8\hat{\gamma} \eta_{\rm C}}\right],
\ee
which is bounded as follows:
\be
\frac{2}{3} \eta_{\rm C} \leq \eta^* \leq \frac{1}{2}
\left[3-\sqrt{9-8\eta_{\rm C}}\right],
\ee
as $\hat{\gamma}$ interpolates in the interval $[0,1]$. These 
bounds were also obtained in Ref. \cite{Holubec2016, VJ2018}.
\section{Refrigerator}
Analogous to the heat engine, one may consider the operation of a refrigerator  
by inverting the thermal and work flows. So, in this case, 
the entropy
generated at the hot and the cold contact is respectively given by:
\be
\Delta S_{{\rm ir},c}^{} =  \Delta S - \frac{Q_c}{T_c},
\label{epc}
\ee
and
\be
\Delta S_{{\rm ir},h}^{} =  \frac{Q_h}{T_h} - \Delta S.
\label{eph}
\ee
Here, $\Delta S>0$ is the entropy change of the working medium
at the cold contact. $Q_c$ is the heat extracted from cold
reservoir, while $Q_h$ is the heat dumped into the hot reservoir.
Within the low-dissipation assumption, the input work to 
drive the refrigerator, $W = Q_h - Q_c$, is given by
\be
W (\gamma, t) = \Delta T \Delta S +  
\left(\frac{\sigma_h}{\gamma} + \frac{\sigma_c}{1-\gamma} \right) \frac{1}{t}.
\label{wgtr}
\ee
As expected, the input work is more than the reversible work,
in case of an irreversible refrigerator.
Then, minimizing the irreversible work w.r.t to $\gamma$, for a given time $t$,
we obtain---as in case of the engine---the optimal value,  
$\hat{\gamma} = {\sqrt{\sigma_h}}/({\sqrt{\sigma_h} + \sqrt{\sigma_c}})$.
So, the optimal input work is given by:
\be
\hat{W}(t) = \Delta T \Delta S + \frac{({\sqrt{\sigma_h} + \sqrt{\sigma_c}})^2}{t},
\label{wmaxtr}
\ee
and the optimal heat extracted from the cold reservoir is
\be
\hat{Q}_c (t) = T_c \Delta S - \frac{\sqrt{\sigma_c}({\sqrt{\sigma_h} + \sqrt{\sigma_c}})}{t}.
\label{qcr}
\ee
The next step would be to obtain an optimal cycle time corresponding to
a chosen objective function, as discussed below.

\subsection{Cooling power}
We consider the cooling power of the refrigerator, operating
with optimal work input for a given cycle time, given by : $\hat{Q}_c (t)/t$. 
The optimal cycle time that maximizes this cooling power is found to be:
\be
t^* = \frac{2  \sqrt{\sigma_c} ({\sqrt{\sigma_h} + \sqrt{\sigma_c}})}
{T_c \Delta S}.
\label{tstarqc}
\ee
The corresponding optimal amounts of heat exchanged with reservoirs are:
\bea
\hat{Q}^{*}_c &=& \frac{T_c \Delta S}{2}, \\
\hat{Q}^{*}_h &=& T_h \Delta S + \sqrt{\frac{\sigma_h}{\sigma_c}} 
\frac{T_c \Delta S}{2}.
\eea
Finally, the coefficient of performance (COP) of the refrigerator
is defined as $\xi = Q_c/(Q_h-Q_c)$, and, 
at optimum cooling power, COP is evaluated to be
\be
\xi^*
= {\xi_{\rm C}} \left(2+\frac{\xi_{\rm C}}{1-\hat{\gamma}}\right)^{-1},
\ee
where $\xi_{\rm C} = T_c/(T_h-T_c)$ is the Carnot coefficient.
The above expression is bounded as 
$0 \leq \xi^*  \leq {\xi_{\rm C}} /(2+{\xi_{\rm C}})$,  
and is also obtained in other studies, 
such as exoreversible refrigerators with only the internal 
irreversibilities \cite{Apertet2013}, and 
within a global linear-irreversible framework for 
total entropy production \cite{Johal2018}, with a parameter
equivalent to $\hat{\gamma}$ and defined in the range $[0,1]$.

It may be noted that a two-parameter, direct optimization problem
cannot be set up with the standard definition of cooling power 
which, from Eq. (\ref{epc}), can be written as:
\be
\frac{Q_c}{t} = \left( T_c  \Delta S - \frac{\sigma_c}{t_c}
\right) \frac{1}{t}.
\label{cpstan}
\ee
Clearly, with $t= t_h + t_c$, the optimum of cooling power
does not exist under the variations of both $t_h$ and $t_c$.

\subsection{Per-unit-time COP}
This objective function was investigated in Ref. \cite{Calvo1997}
for the endoreversible model. It is a criterion
for refrigerators, analogous
to the function used in Section IIIB on engines.
Again, we use the optimal work condition for a given cycle time,
and evaluate $\hat{\xi}(t) \equiv \hat{Q}_c/ \hat{W}$, using 
Eqs. (\ref{wmaxtr}) and (\ref{qcr}). Then, we optimize 
the function:
\be
\dot{\hat{\xi}} \equiv \frac{\hat{\xi}(t)}{t},
\ee
w.r.t time $t$. The COP at the optimal $\dot{\hat{\xi}}$ is evaluated
to be:
\be
\hat{\xi}^* =  (1-\hat{\gamma})\left[\sqrt{ 1 + \frac{\xi_{\rm C}}{1-\hat{\gamma}}}-1 \right].
\ee
which is bounded as: $ 0 \leq \hat{\xi}^* \leq \sqrt{ 1+  \xi_{\rm C}} -1$.
The bounds match with the findings of Ref. \cite{Calvo1997}.
\subsection{$\chi$-criterion}
In the literature on the optimal performance
of refrigerators, $\chi$-criterion is defined as: $\chi = \xi Q_c/t$.
This has been studied within endoreversible \cite{Chen1990}  as well as
low-dissipation models \cite{Roco2012, Roco2012b}. 
As pointed out above, the calculations
may become involved for such objective functions, 
thus making the analytic solution intractable \cite{Roco2012}.
However, the present approach of two-step optimization
leads directly to an exact expression
for the COP as:
\be
\xi^* = \frac{1-\hat{\gamma}}{2} 
\left[ \sqrt{9 + \frac{8 \xi_{\rm C}}{1-\hat{\gamma}}} -3 \right].
\ee
Interestingly, a similar formula as above is obtained
for the so-called minimally nonlinear irreversible refrigerators 
\cite{Roco2013}, with an
equivalent parameter defined in the range [0,1]. 
Moreover, the above formula satisfies the 
following bounds:
\be
0 \leq \xi^* \leq \frac{1}{2}\left[ 
\sqrt{9 + {8 \xi_{\rm C}}} -3 \right],
\ee
which are obtainable under the standard optimization \cite{Roco2012}.

\section{Discussion}
In this section, we further highlight some  
features of the low-dissipation model, in particular, those
related to the first step of work optimization for a given cycle time.
We start by studying the notion of lost work in this model. 

The lost work is equivalent to the energy which is made unavailable for work, due
to the irreversible process \cite{Tolman1948}.
As a way to characterize irreversibility in nonequilibrium thermodynamics, it can 
also be related to the concept of thermodynamic
length in finite-time thermodynamics \cite{Salamon1983}.
It is defined by the difference:
$W_{\rm lost} = W_{\rm rev} - W$.
Therefore, optimizing the
work in the irreversibile process 
as above, implies minimizing the lost work.
Since, $W_{\rm rev} = \Delta T \Delta S$, from Eq. (\ref{wmaxt}), we have
\be
\hat{W}_{\rm lost} = \frac{({\sqrt{\sigma_h} + \sqrt{\sigma_c}})^2}{t},
\label{wlost}
\ee
as the minimum lost work in the low-dissipation engine with a given cycle 
time $t$. Further, according to Gouy-Stodola theorem \cite{Gouy, Stodola, Tolman1948},
the lost work is directly proportional to the total entropy produced in the 
irreversibile process, where the constant of proportionality 
is given by a reference temperature $T_0$. 
Then, the above theorem as applied at optimal work, implies:
\be
\hat{W}_{\rm lost}  = T_0 \Delta _{\rm tot} \hat{S}.
\label{gs}
\ee 
Now, the total entropy produced per cycle
(of a given time $t$) is the sum of entropies produced
at the hot and cold steps. Under optimal work:
\bea
\Delta _{\rm tot} \hat{S} 
& =& \Delta _{\rm ir} \hat{S}_h + \Delta _{\rm ir} \hat{S}_c, \\
&=& \frac{\sigma_h}{\hat{\gamma} T_h t} 
+ \frac{\sigma_c}{(1-\hat{\gamma}) T_c t}, \nonumber \\
& = &  
\left[ \frac{\sqrt{\sigma_h}}{T_h} + \frac{\sqrt{\sigma_c}}{T_c}  \right]
\frac{{\sqrt{\sigma_h} + \sqrt{\sigma_c}}}{t}, \nonumber \\
& = &  
\left[ \frac{\hat{\gamma}}{T_h} + \frac{1-\hat{\gamma}}{T_c}  \right]
\frac{({\sqrt{\sigma_h} + \sqrt{\sigma_c}})^2}{t}.
\label{stotmin}
\eea
Comparing Eq. (\ref{wlost}) with (\ref{stotmin}), we obtain: 
\be
{T}_0 = \left[ \frac{\hat{\gamma}}{T_h} + \frac{1-\hat{\gamma}}{T_c}   \right]^{-1},
\label{thm}
\ee
a weighted harmonic mean of the reservoir temperatures.
Clearly, $T_c \leq T_0 \leq T_h$.
It may be noted that the apparent form of ${T}_0$ as a harmonic mean,
holds for the case when the dissipation constants $\sigma_j$, and hence
the weights $\{\hat{\gamma}, 1-\hat{\gamma}\}$, are 
temperature-independent. The specific form of $T_0$
is different if $\hat{\gamma}$ depends on temperatures.
For instance,  based on a general, microscopic approach \cite{Cavina2017},
the dissipation 
constants are found to obey: ${\sigma_c/ \sigma_h} = (T_c/T_h)^{2\alpha}$,
where $\alpha \in R$. Thereupon, $T_0$ comes out in the form:
\be
T_0 = 
\dfrac{T_{h}^{\alpha} +T_{c}^{\alpha} }{T_{h}^{\alpha-1}+ T_{c}^{\alpha-1}},
\label{lehmer}
\ee
which is known by the name of Lehmer mean \cite{Bullen2003}. 
Clearly, some standard means are subsumed in 
this general case. In particular, for 
$\alpha=0$, we obtain the symmetric harmonic mean.
For $\alpha =1/2$,
we obtain the geometric mean, $T_0 = \sqrt{T_h T_c}$,
and the arithmetic mean, for $\alpha =1$.

Finally, it is interesting to rewrite Eq. (\ref{stotmin}), as follows:
\be
T_0 \Delta _{\rm tot} \hat{S} = \frac{({\sqrt{\sigma_h} + \sqrt{\sigma_c}})^2}{t}.
\label{dstotld}
\ee
Thus, we note that if the lost work is minimized for a fixed cycle time, 
it yields a concrete expression for the reference temperature 
relating the lost work to the total entropy produced 
(Gouy-Stodola theorem). Secondly, the total entropy
produced is rendered inversely proportional to
the cycle time, which is reminiscent of the low-dissipation
behavior, but now applied to the cycle as a whole. Consequently, 
one may as well identify an effective dissipation
constant for the total cycle, in Eq. (\ref{dstotld}), as 
$({\sqrt{\sigma_h} + \sqrt{\sigma_c}})^2$. In this sense,
the first step of optimizing work may be seen as a natural
step, since it extends the low-dissipation behavior 
from the individual processes to the overall cycle.
\section{Conclusions}
We have proposed and demonstrated the utility of an alternate,
two-step optimization procedure for low-dissipation thermal machines,
whereby the irreversible work is optimized first for a given cycle time,
and then a second objective function is optimized,
yielding the optimal cycle time for the operation of the device.
It is important to emphasize that, in general, the optimum
with the alternate procedure will not coincide with the 
global optimum of the chosen objective function 
in the standard approach (simultaneous variation
of the two contact times), except in case of power output. 
Actually, the standard approach may not yield a tractable 
solution for the global optimization problem, such as
$\chi$-criterion for low-dissipation refrigerators \cite{Roco2012}.
Further, a two-parameter optimization problem may not
be well-defined for some objective function, e.g. cooling
power of low-dissipation refrigerator.
On the other hand, the present approach yields, 
rather easily, closed-form expressions 
as well as bounds of figures of merit for such functions. 
Interestingly, these bounds match with the corresponding bounds 
 (whenever these can be derived) from the standard
approach.
We have applied this approach to a few objective 
functions, such as efficient power, per-unit-time efficiency,
and $\chi$-criterion, which are not easy to treat within the standard 
approach. 

Further, it is observed for various objective functions
that the optimal results come out  equivalent
to those obtained from other models, such as
the endoreversible, minimally nonlinear irreversible approaches
and global linear-irreversibile framework.
This analogy between the low-dissipation and 
the endoreversible models \cite{Johal2017} as well as 
other approaches \cite{Izumida_2012, Roco2012, Johal2018}
needs to be further explored.  One reason is that 
the constraints for optimization, such as keeping $\Delta S$ fixed 
\cite{Chen1998}, are the same in some procedures.
It is hoped 
that the proposed scheme will provide insights into the connections
between optimal behavior of different irreversible models.  
Finally, it is straightforward to extend the present analysis 
to a multi-reservoirs scenario \cite{Broeck2013}. 
\acknowledgments
{The author gratefullly acknowledges useful discussions with 
Viktor Holubec, I. Iyyappan, Steffen Rulands and Varinder Singh.}

%

\end{document}